\documentclass{elsart}
\usepackage{amsmath}

\DeclareMathOperator{\im}{Im}
\DeclareMathOperator{\tr}{tr}
\DeclareMathOperator{\trg}{trg}

\DeclareMathOperator{\diag}{diag}

\def\openone{\leavevmode\hbox{\small1\kern-3.3pt\normalsize1}}

\begin{document}

\begin{frontmatter}

\title{Spectral correlations of the massive QCD Dirac operator at
  finite temperature} 
\author[Muenchen]{Burkhard Seif},
\author[Muenchen]{Tilo Wettig},
\author[Heidelberg]{Thomas Guhr}
\address[Muenchen]{Institut f\"ur Theoretische Physik,
  Technische Universit\"at M\"unchen, D-85747 Garching, Germany}
\address[Heidelberg]{Max-Planck-Institut f\"ur Kernphysik, 
         Postfach 103980, D-69029 Heidelberg, Germany}

\date{17 February 1999}

\begin{abstract}
  We use the graded eigenvalue method, a variant of the supersymmetry
  technique, to compute the universal spectral correlations of the QCD
  Dirac operator in the presence of massive dynamical quarks.  The
  calculation is done for the chiral Gaussian unitary ensemble of
  random matrix theory with an arbitrary Hermitian matrix added to the
  Dirac matrix.  This case is of interest for schematic models of
  QCD at finite temperature.\\[4mm]
\end{abstract}

\end{frontmatter}

\section{Introduction}
\label{intro}

By now, it has been firmly established that the spectrum of the QCD
Dirac operator possesses a number of universal features which can be
described by chiral random matrix theory (RMT) \cite{Leut92,Shur93}.
In particular, the RMT predictions agree very well with data from
lattice gauge simulations, both for the eigenvalue correlations in the
bulk of the spectrum on the scale of the mean level spacing
\cite{Hala95,Mark98,guhr98} and for the distribution and correlations
of the low-lying eigenvalues \cite{Berb98a,Ma98,Berb98b,Damg98}.  It
is the key assumption of chiral RMT that the matrix elements of the
Dirac operator in an appropriate energy basis behave as random
numbers.  This concept has already been very successful in many other
areas of physics, see the detailed review in Ref.~\cite{review}. Thus,
it is fair to say that RMT approaches can be viewed as thermodynamics
for spectral fluctuations and related properties. In the context of
QCD, deviations from pure RMT statistics have also been found in the
microscopic \cite{osborn98,bitsch98} and in the bulk region
\cite{guhr98}.  These findings provide evidence for the conjecture
that lattice QCD may have much in common with disordered systems
\cite{stern98,janik98,osborn98}. In the present work, however, we
focus on some formal and theoretical aspects of chiral RMT.  Hence, we
shall not discuss in any detail the physical applications but refer to
the existing literature \cite{Verb97,Berb98c}.

Our aim here is to compute the universal spectral correlations of the
QCD Dirac operator in the presence of massive dynamical quarks for the
chiral Gaussian unitary ensemble of RMT.  Such a calculation has been
done previously at zero temperature using orthogonal polynomials
\cite{Damg97a,Wilk98} and the finite volume partition function
\cite{Akem98a,Akem98b}.  In Ref.~\cite{Toub98}, the connection between
these two results was established in the framework of partially
quenched chiral perturbation theory.  In this paper, we employ the
graded eigenvalue method \cite{Guhr91,guhr96a,guhr96b}, which is a
special variant of the supersymmetry method \cite{Efet83,Verb85}, for
the following two reasons.  First, it allows us to extent the
calculation to the case where an arbitrary deterministic Hermitian
matrix is added to the Dirac matrix.  This is relevant for schematic
random-matrix models of QCD at finite temperature
\cite{Jack95,Wett96}.  The standard orthogonal-polynomial method
cannot easily be applied in this case, because a certain rotation
invariance in the space of the random matrices is lost.  Second, the
present problem leads to an interesting extension of the graded
eigenvalue method in the context of chiral RMT. In ordinary RMT, this
method was developed to calculate spectral correlations in crossover
transitions from regularity to chaos \cite{guhr96a,guhr96b}.  This
method was then extended to chiral RMT and used to compute the
universal spectral correlations of the Dirac operator in the quenched
approximation, i.e., without dynamical quarks \cite{Guhr97a,Jack97a}.
For an application of the supersymmetry method to the case of one
flavor at zero temperature, see Ref.~\cite{Jurk96}.

In Refs.~\cite{Damg97a,Split98} it was shown that the RMT results are
invariant under deformations of the distribution of the random matrix.
While we suspect this statement to hold also for the results computed
in the present paper, a rigorous proof would require an extension of
the work of Ref.~\cite{Zinn98}.  We shall not address this issue here.

The outline of this paper is as follows.  In Sec.~\ref{setup}, we
define the problem and outline the main idea for the solution.  The
graded eigenvalue method is applied in Sec.~\ref{susy}.  The special
case of energies and masses on the scale of the mean level spacing
near zero, which is also the most interesting case for physical
applications, is considered in Sec.~\ref{micro}.  We conclude with a
summary in Sec.~\ref{summary}.  Technical details of the calculation
are discussed in two appendixes.

\section{Setup of the calculation}
\label{setup}

Since this paper is a natural extension of Ref.~\cite{Guhr97a}, we
shall attempt to use a similar notation.  The QCD Dirac operator in
Euclidean space is defined by $D=\gamma_\mu\partial_\mu+ig\gamma_\mu
A_\mu$, where $g$ is the coupling constant and the $A_\mu$ are the
gauge fields.  Note that $D$ is anti-Hermitian.  In a random matrix
model in a chiral basis, the matrix $A$ representing the Dirac
operator has the form \cite{Shur93}
\begin{equation}
  \label{eq1.1}
  D\longrightarrow iA=
  i\begin{bmatrix}0 & W+Y \\ W^\dagger+Y & 0\end{bmatrix}\:,
\end{equation}
where $W$ is a square random matrix of dimension $N$ and $Y$ is an
arbitrary Hermitian matrix.  Expression~\eqref{eq1.1} is a {\em
  schematic\/} model for the QCD Dirac operator at finite temperature.
The matrix $Y$ represents the effects of the temperature on the Dirac
spectrum.  Its specific form depends on the choice of basis states
\cite{Jack95,Wett96}.  Since we consider an arbitrary Hermitian matrix
$Y$, we cover all possible choices of basis states.

One could also consider the more general problem of a rectangular
matrix $W$, giving rise to exact zero modes of the Dirac operator
whose number can be identified with the topological charge.  At $Y=0$,
this is not necessary since this problem is equivalent to introducing
additional massless flavors \cite{Verb94a}.  Although it is not
obvious, we expect this equivalence to hold also for nonzero $Y$.  We
hope to address this problem in future work.

We will be interested in the correlations of the eigenvalues of the
matrix $A$.  In this paper, we study the chiral Gaussian unitary
ensemble (chGUE) appropriate for QCD with three or more colors for
which $W$ is a complex matrix without any symmetries \cite{Verb94a}.
The probability distribution of $W$ is given by
\begin{align}
  \label{eq1.2}
  P(W)&=\frac{1}{\mathcal{N}}\,P_0(W)\prod_{f=1}^{N_f} \det(im_f-A)\:,
\intertext{with}
  \label{eq1.2a}
  P_0(W)&=\exp\left(-N\Sigma^2\tr WW^\dagger\right)\:.
\end{align}
Here, $N_f$ is the number of quark flavors with masses $m_f$,
$\mathcal{N}$ is a normalization factor (see below), and $\Sigma$ is a
real parameter which will turn out to be equal to the chiral
condensate at zero temperature.  Note that $\mathcal{N}$ depends on
the quark masses.  To be precise, the argument of the determinant in
\eqref{eq1.2} should have been $m_f+D=m_f+iA$.  For convenience, we
have pulled out a factor of $-i$ and absorbed it in $\mathcal{N}$.

We are interested in the $k$-point spectral correlation functions,
defined as the probability of finding energies in infinitesimal
intervals around the points $x_1,\dots,x_k$, regardless of labeling.
Apart from some trivial contributions involving $\delta$-functions
\cite{review} they are given by
\begin{equation}
  \label{eq1.3}
  R_k(x_1,\dots,x_k)=\left(-\frac{1}{\pi}\right)^k \int d[W] P(W)
  \prod_{p=1}^k \im \tr \frac{1}{x_p^+-A} \:,
\end{equation}
where $x_p^+=x_p+i\varepsilon$ with $\varepsilon$ positive
infinitesimal.  The measure $d[W]$ is simply the product of the
differentials of the real and imaginary parts of the elements of $W$,
i.e., of all independent variables.  The integrations extend from
$-\infty$ to $+\infty$.  The normalization factor $\mathcal{N}$ is
determined by the requirement $\int d[W]P(W)=1$.  Since both the
distribution~(\ref{eq1.2}) and the measure $d[W]$ are invariant under
unitary transformations of $W$, only the relative unitary rotation
between $W$ and $Y$ matters and we can, without loss of generality,
write the Hermitian matrix $Y$ in diagonal form,
$Y=\diag(y_1,\dots,y_N)$.  Advantageously, the $k$-point functions can
be obtained as
\begin{equation}
  \label{eq1.6}
  R_k(x_1,\dots,x_k)=\left(-\frac{1}{\pi}\right)^k
  \left.\frac{\partial^k}{\prod_{p=1}^k \partial J_p}Z_k(J) 
  \right|_{J_p=0} 
\end{equation}
with a generating function given by
\begin{equation}
  \label{eq1.5}
  Z_k(J) = \int d[W] P(W) \prod_{p=1}^k\det(x_p-A)
  \,\im\frac{1}{\det(x_p^+-J_p-A)} \:,
\end{equation}
where $J$ stands for $J_1,\dots,J_k$.  The starting point of the
graded eigenvalue method is to rewrite the determinants in
\eqref{eq1.5} as Gaussian integrals over commuting and anti-commuting
variables.  Note that the distribution $P(W)$ contains $N_f$
determinants in the numerator.  To retain the determinant structure of
the problem, it is highly desirable to have an equal number of
determinants in numerator and denominator so that the bosonic and
fermionic blocks in the supersymmetric representation of the
generating function have the same size.  Therefore, we introduce $N_f$
additional determinants in the denominator and write, in the large-$N$
limit,
\begin{equation}
  \label{eq1.7a}
  Z_k(J)=\lim_{\{a_f\}\to\infty}\tilde Z_k(J)
\end{equation}
with
\begin{align}
  \label{eq1.7}
  \tilde Z_k(J)= \frac{1}{\mathcal{\tilde{N}}}\int d[W] \, P_0(W) &
  \prod_{p=1}^k\det(x_p-A)\,\im\frac{1}{\det(x_p^+-J_p-A)}\nonumber\\
  \times&\prod_{f=1}^{N_f}\det(im_f-A)\,\im\frac{1}{\det(a_f^+-A)}\:,
\end{align}
where the $a_f$ ($f=1,\dots,N_f$) are dummy real variables.  The
modified normalization $\mathcal{\tilde{N}}$ depends on the $a_f$ and
is given by
\begin{equation}
 \label{eq1.8}
 \tilde{\mathcal{N}} = \int d[W] \, P_0(W) \prod_{f=1}^{N_f}
 \det(im_f-A)\,\im \frac{1}{\det(a_f^+-A)}\:.
\end{equation}
The fact that Eq.~\eqref{eq1.7a} holds in the limit $N\to\infty$ is
proved in App.~\ref{app:Zktilde}.  The introduction of the dummy
determinants in \eqref{eq1.7} is the main idea of the present
calculation.  As we shall see below, it allows us to use the results
of Ref.~\cite{Guhr97a} so that the generalization from the quenched
approximation to the case with $N_f>0$ can be obtained with moderate
effort.

We observe that \eqref{eq1.8} is essentially a special case of
\eqref{eq1.7} with $k=0$.  To simplify the notation, we will compute
the generic quantity
\begin{equation}
  \label{generic}
  G_\gamma(t)=\int d[W] \, P_0(W)\prod_{j=1}^\gamma\det(t_{j2}-A)
  \,\im\frac{1}{\det(t_{j1}^+-A)}\:,
\end{equation}
where $\gamma$ is a nonnegative integer and
$t=\diag(t_{11},\dots,t_{\gamma 1},t_{11},\dots,t_{\gamma 2})$ is a
diagonal graded matrix of dimension $2\gamma$.  Both \eqref{eq1.7} and
\eqref{eq1.8} can be obtained from \eqref{generic} by choosing
$\gamma$ and $t$ appropriately.

\section{Supersymmetric representation and graded eigenvalue method}
\label{susy}

Since we can employ the results of Ref.~\cite{Guhr97a} to compute the
function \eqref{generic} very efficiently, we only review the major
steps in the derivation.  First, the determinants in \eqref{generic}
are rewritten as Gaussian integrals over commuting and anti-commuting
variables which are arranged in a graded (or super) vector $\psi$.
Then, the integration over $W$ can be performed, resulting in
fourth-order terms in the $\psi$-variables.  These terms can be
removed by a Hubbard-Stratonovitch transformation at the expense of
introducing additional integration variables which can be arranged in
a complex graded (or super) matrix $\sigma$.  The order of the
integrations over $\sigma$ and $\psi$ can then be interchanged,
provided that one is only interested in the imaginary parts in
\eqref{generic}.  This point has been discussed in
Refs.~\cite{Guhr97a,Jack97a,Jack96b}.  The $\psi$-integration can then
be performed trivially.  The graded matrix $\sigma$ can be written in
spherical coordinates as $us\bar v$ with graded (or super) unitary
matrices $u$ and $\bar v$ and radial coordinates $s$.  The integration
over $u$ and $\bar v$ can be performed using the supersymmetric
generalization of the Berezin-Karpelevich integral \cite{Guhr96}. It
can be viewed as the extension of the supersymmetric Itzykson-Zuber
integral \cite{Guhr91} to complex graded matrices.  The final result
for the function \eqref{generic} then becomes (see Eqs.~(33) through
(36) of Ref.~\cite{Guhr97a})
\begin{equation}
  \label{eq2.1}
  G_\gamma(t)=\left(\frac{\pi}{N\Sigma^2}\right)^{N^2}
  \frac{\exp(-N\Sigma^2\trg t^2)}{B_\gamma(t^2)} 
  \det[C_N(t_{i1},t_{j2})]_{i,j=1,\dots,\gamma}
\end{equation}
with 
\begin{align}
  \label{eq2.2}
  C_N(x_1,x_2)=(2N\Sigma^2&)^2
  \int\limits_0^\infty\int\limits_0^\infty  ds_1ds_2
  \frac{s_1s_2}{s_1^2+s_2^2}\exp\left(-N\Sigma^2(s_1^2+s_2^2)\right)
  \nonumber\\
  &\times I_0(2N\Sigma^2s_1x_1)J_0(2N\Sigma^2s_2x_2)\im\prod_{n=1}^N
  \frac{y_n^2+s_2^2}{y_n^2-(s_1^+)^2}\:,
\end{align}
where $J$ and $I$ denote the Bessel and modified Bessel function,
respectively.  In Eq.~\eqref{eq2.1}, the symbol trg denotes the graded
trace, and 
\begin{equation}
  \label{berezinian}
  B_\gamma(t^2)=\frac{\Delta_\gamma(t_1^2)\Delta_\gamma(t_2^2)}
  {\prod_{ij}(t_{i1}^2-t_{j2}^2)}
  \qquad\text{with}\qquad
  \Delta_\gamma(x)=\prod_{i>j}^\gamma(x_i-x_j)\:.
\end{equation}
Equations~\eqref{eq1.7} and \eqref{eq1.8} can now be obtained by
choosing $\gamma=k+N_f$ and $\gamma=N_f$, respectively, and
substituting appropriate values for the entries of $t$.  Performing
the differentiations according to Eq.~\eqref{eq1.6} (with $Z_k(J)$
replaced by $\tilde Z_k(J)$, see Eq.~\eqref{eq1.7a}) then yields the
$k$-point functions.  We obtain after some algebra
\begin{equation}
  \label{eq2.5}
  R_k(x_1, \dots, x_k) = \left(\frac2\pi\right)^k
  \left(\prod_{p=1}^kx_p\right)
  \lim_{\{a_f\}\to\infty}
  \frac{\det[C_N(z_p,\zeta_q)]_{p,q=1,\dots,k+N_f}}
  {\det[C_N(a_f,im_g)]_{f,g=1,\dots,N_f}}
\end{equation}
with
\begin{align}
  \label{z_p}
  z_p=&
  \begin{cases}
    x_p\phantom{ww}&     \text{for $p=1, \dots, k$,}\\
    a_{p-k}& \text{for $p=k+1, \dots, k+N_f$,}
  \end{cases}
  \\[3mm]
  \label{zeta_q}
  \zeta_q=&
  \begin{cases}
    x_q\phantom{ww}&      \text{for $q=1, \dots, k$,}\\
    im_{q-k}& \text{for $q=k+1, \dots, k+N_f$.}
  \end{cases}
\end{align}
Note that Eq.~\eqref{eq2.5}, just like Eq.~\eqref{eq1.7a}, holds in
the limit $N\to\infty$ (which is the interesting limit for physical
applications) but not for finite $N$.  Instead of introducing dummy
determinants in Eq.~\eqref{eq1.7}, there is an alternative way to
proceed which is exact for finite $N$.  We briefly explain the idea
for readers familiar with the graded eigenvalue method.  Without the
introduction of the dummy determinants in Eq.~\eqref{eq1.7}, the
transformation from ordinary space to superspace leads to a graded (or
super) matrix $\sigma$ whose boson-boson and fermion-fermion blocks
have dimension $k$ and \mbox{$k+N_f$}, respectively.  The
transformation of $\sigma$ to spherical coordinates involves a
Berezinian which cannot be written as a determinant.  The idea now is
to enlarge the boson-boson block of $\sigma$ to dimension $k+N_f$ by
introducing dummy integration variables in superspace.  Then, the
Berezinian resulting from the enlarged $\sigma$-matrix can be written
as a determinant which is the prerequisite for expressing the
$k$-point function in form of a determinant.  However, we will not
discuss this alternative way since we are only interested in the limit
$N\to\infty$ for which the method of Eqs.~\eqref{eq1.7a} through
\eqref{eq1.8} appears to be more economic.

Our conventions are such that the support of the spectral density is
of order $\mathcal{O}(1)$ and the typical level spacing is of order
$\mathcal{O}(1/N)$.  While Eq.~\eqref{eq2.5} holds for all values of
the $x_p$ and $m_f$, we are particularly interested in the microscopic
region where the $x_p$ and $m_f$ are of order $\mathcal{O}(1/N)$.  We
now turn to this limit.

\section{Microscopic limit}
\label{micro}

If $x_1$ and $x_2$ in Eq.~\eqref{eq2.2} are of order
$\mathcal{O}(1/N)$, the integrals can be performed in saddle-point
approximation in the large-$N$ limit.  This was done in
Ref.~\cite{Guhr97a}, and we obtain the Bessel kernel (see Eq.~(63) of
\cite{Guhr97a})
\begin{equation}
  \label{eq3.1}
  C_N(x_1,x_2)= \pi N\Xi\frac{x_1J_1(2N\Xi x_1)J_0(2N\Xi x_2)-
    x_2J_0(2N\Xi x_1)J_1(2N\Xi x_2)}{x_1^2-x_2^2}\;,
\end{equation}
where $\Xi$ is the only real and positive solution of
\begin{equation}
  1=\frac{1}{N}\sum_{n=1}^N\frac{1}{(\Sigma y_n)^2+(\Xi/\Sigma)^2}
\end{equation}
or zero if no such solution exists \cite{Guhr97a}.  As we shall show
below, $\Xi=\Xi(Y)$ can be identified with the chiral condensate in
the presence of the arbitrary offset $Y$.  We now rescale the
energies, the masses, and the dummy variables by $2N\Xi$ and define
$u_p=2N\Xi x_p$, $\mu_f=2N\Xi m_f$, and $\alpha_f=2N\Xi a_f$.  We thus
obtain from \eqref{eq2.5}
\begin{equation}
  \label{eq3.3}
  R_k(x_1,\dots,x_k) = (2N\Xi)^k\left(\prod_{p=1}^ku_p\right)
  \lim_{\{\alpha_f\}\to\infty}
  \frac{\det[C(\tilde z_p,\tilde \zeta_q)]_{p,q=1,\dots,k+N_f}}
  {\det[C(\alpha_f,i\mu_g)]_{f,g=1,\dots,N_f}}  
\end{equation}
with a kernel given by
\begin{equation}
  \label{kernel}
  C(u_1,u_2)=\frac{u_1J_1(u_1)J_0(u_2)-u_2J_0(u_1)J_1(u_2)}
  {u_1^2-u_2^2}\:,
\end{equation}
where we have used the notation $\tilde z_p=2N\Xi z_p$ and
$\tilde\zeta_p=2N\Xi\zeta_p$.  We shall also need this kernel for the
second argument purely imaginary,
\begin{equation}
  \label{ikernel}
  C(u,i\mu)=\frac{uJ_1(u)I_0(\mu)+\mu J_0(u)I_1(\mu)}{u^2+\mu^2}\:,
\end{equation}
and in the special case $u_1=u_2=u$,
\begin{equation}
  \label{ukernel}
  C(u,u)=\frac12\left[J_0^2(u)+J_1^2(u)\right]\:.
\end{equation}

The major difficulty now is to perform the limit
$\{\alpha_f\}\to\infty$ in \eqref{eq3.3}.  For this purpose, it is
convenient to divide both numerator and denominator of the last term
in \eqref{eq3.3} by $\Delta_{N_f}(\alpha)$ and to perform the limits
separately in numerator and denominator.  Since the derivation is
somewhat technical it is presented in App.~\ref{app:limit}.  The final
result is
\begin{equation}
  \label{eq3.5}
  \lim_{\{\alpha_f\}\to\infty}
  \frac{\det[C(\tilde z_p,\tilde \zeta_q)]_{p,q=1,\dots,k+N_f}}
  {\det[C(\alpha_f,i\mu_g)]_{f,g=1,\dots,N_f}}=
  \frac{\det[\mathcal{A}_{pq}]_{p,q=1,\dots,k+N_f}}
  {\det[\mathcal{B}_{fg}]_{f,g=1,\dots,N_f}}\:,
\end{equation}
where
\begin{equation}
  \label{Bmatrix}
  \mathcal{B}=
  \begin{bmatrix}
    I_0(\mu_1)&\cdots&I_0(\mu_{N_f})\\
    -\mu_1I_1(\mu_1)&\cdots&-\mu_{N_f}I_1(\mu_{N_f})\\
    \vdots&&\vdots\\
    (-\mu_1)^{N_f-1}I_{N_f-1}(\mu_1)&\cdots&
    (-\mu_{N_f})^{N_f-1}I_{N_f-1}(\mu_{N_f})
  \end{bmatrix}
\end{equation}
and
\begin{equation}
  \label{Amatrix}
  \mathcal{A}=
  \begin{bmatrix}
    C(u_1,u_1)&\cdots&C(u_1,u_k)&
    C(u_1,i\mu_1)&\!\cdots\!&C(u_1,i\mu_{N_f})\\
    \vdots&&\vdots&\vdots&&\vdots&\\
    C(u_k,u_1)&\cdots&C(u_k,u_k)&
    C(u_k,i\mu_1)&\!\cdots\!&C(u_k,i\mu_{N_f})\\
    J_0(u_1)&\cdots&J_0(u_k)&&&\\
    u_1J_1(u_1)&\cdots&u_kJ_1(u_k)&&&\\
    \vdots&&\vdots&&\mathcal{B}&\\
    u_1^{N_f-1}J_{N_f-1}(u_1)&\cdots\!&u_k^{N_f-1}J_{N_f-1}(u_k)&&&
  \end{bmatrix}
\end{equation}
In compact notation, we have
\begin{equation}
  \label{A}
  \mathcal{A}_{pq}=
  \begin{cases}
    C(u_p,u_q)&\text{for $1\le p,q\le k$};\\
    C(u_p,i\mu_{q-k})&\text{for $1\le p\le k$; $k+1\le q\le k+N_f$};\\
    u_q^{p-k-1}J_{p-k-1}(u_q)&
    \text{for $k+1\le p\le k+N_f$; $1\le q\le k$};\\
    (-\mu_{q-k})^{p-k-1}I_{p-k-1}(\mu_{q-k})&
    \text{for $k+1\le p,q\le k+N_f$}
  \end{cases}
\end{equation}
and
\begin{equation}
  \label{B}
  \mathcal{B}_{fg}=(-\mu_g)^{f-1} I_{f-1}(\mu_g) \qquad 
  \text{for $1\le f,g\le N_f$}\:.
\end{equation}
Rescaling the $R_k$ by $(2N\Xi)^{-k}$, we arrive at the final result
for the microscopic spectral correlations,
\begin{align}
  \label{final}
  \rho_k(u_1,\dots,u_k)&
  \equiv\frac{1}{(2N\Xi)^k}R_k(x_1,\dots,x_k)\nonumber\\
  &=\left(\prod_{p=1}^ku_p\right)
  \frac{\det[\mathcal{A}_{pq}]_{p,q=1,\dots,k+N_f}}
  {\det[\mathcal{B}_{fg}]_{f,g=1,\dots,N_f}}\:.
\end{align}
It remains to be shown that $\Xi$ can be identified with the absolute
value of the chiral condensate $\langle\bar\psi\psi\rangle$.
According to the Banks-Casher relation \cite{Bank80}, we have
$V|\langle\bar\psi\psi\rangle|=\pi R_1(0)$, where $R_1(0)$ is the
spectral density at zero.  The space-time volume $V$ can be identified
with $2N$.  Furthermore, we have
\begin{equation}
  R_1(0)=2N\Xi\lim_{u\to\infty}\rho_1(u)\:.
\end{equation}
To compute this limit, we use the matrix $\mathcal{A}$ in
\eqref{Amatrix} with $k=1$ and $u_1=u$.  We first observe that
$C(u,u)\to1/(\pi u)$ as $u\to\infty$, see \eqref{ukernel}.  In
addition, the denominator of the entries $C(u,i\mu_f)$, see
\eqref{ikernel}, can be written as a geometric series in $1/u^2$ so
that $C(u,i\mu_f)$ is given as an expansion in $1/u^2$ with
coefficients containing $I_0(\mu_f)$, $\mu_fI_1(\mu_f)$, etc.  By
subtracting appropriate multiples of rows 2 to $N_f+1$, the first
$N_f$ of these terms can be eliminated, see also the discussion of
Eq.~\eqref{app1.2} in App.~\ref{app:limit}.  The leading large-$u$
behavior of $C(u,u)$ is not modified by these subtractions.
Higher-order terms in the expansion of $C(u,i\mu_f)$ are suppressed by
powers of $1/u$, the leading term being of order
$J_{(N_f+1)\,\text{mod}\,2}(u)/u^{N_f+1}$.  Even when multiplied by
the largest entry in the lower-left corner of $\mathcal{A}$,
$u^{N_f-1}J_{N_f-1}(u)$, the result is suppressed compared to $1/(\pi
u)$.  In the large-$u$ limit, the determinant of \eqref{Amatrix} with
$k=1$ thus becomes $1/(\pi u)\cdot\det\mathcal{B}$, and we obtain
\begin{equation}
  2N\Xi\lim_{u\to\infty}\rho_1(u)=2N\Xi u \frac{1}{\pi u}
  =\frac1\pi 2N\Xi
\end{equation}
and, hence, $2N\Xi=\pi R_1(0)$ as desired.  Therefore, we have shown
that the functional form of the microscopic spectral correlations in
the presence of massive dynamical quarks does not change if a
deterministic matrix $Y$ is added to the matrix of the Dirac operator,
provided that $\Xi$, i.e., the chiral condensate, is nonzero.  The
only dependence of the final result on the matrix $Y$ appears in form
of a rescaling of the energy scale, from $\Sigma$ at $Y=0$ to $\Xi$ at
$Y\ne0$.

Our final result \eqref{final} is given in terms of the determinant of
a $(k+N_f)\times(k+N_f)$ matrix whose entries are simple functions.
This structure arises naturally in the graded eigenvalue method.  Two
other forms for the microscopic spectral correlations have been obtain
previously.  In the orthogonal-polynomial method, the result is given
as the determinant of a $k\times k$ matrix whose entries are
$(N_f+2)\times(N_f+2)$ matrices \cite{Damg97a,Wilk98}.  {}From the
finite volume partition function, the result is given as the
determinant of a $(2k+N_f)\times(2k+N_f)$ matrix whose entries are
simple functions \cite{Akem98b}.  At the present time, we do not have
a closed mathematical proof that these three results are identical.
However, we have perfomed extensive checks for a large number of
different values of $k$ and $N_f$, both numerically and using computer
algebra.  In all cases, the three results agree perfectly so that we
do not have any doubt that they are identical.  One of the virtues of
the present method is that it allows us to include the deterministic
matrix $Y$ as well.  Moreover, it appears to lead to the most
economical representation of the final result.

In Ref.~\cite{Wilk98}, the distribution of the smallest eigenvalue,
$P(\lambda_{\rm min})$, was also computed.  Its universality with
respect to deformations of $P_0(W)$ was shown in Ref.~\cite{Nish98}.
Since $P(\lambda_{\rm min})$ follows directly from the microscopic
spectral density $\rho_1(u)$ \cite{Wilk98}, and since we have shown in
this paper that the functional form of the latter quantity is not
affected by the addition of $Y$, it follows that the functional form
of $P(\lambda_{\rm min})$ also remains unchanged.

\section{Summary}
\label{summary}

In this paper, we have extended the graded eigenvalue method to the
case where massive dynamical quarks enter the distribution of the
random matrix.  The virtue of this approach is twofold.  First, we have
obtained a novel representation of results computed previously with
other methods.  Our representation appears to be the most economical
one.  It is also very stable numerically, compared to the
representations obtained from the orthogonal-polynomial method and the
finite-volume partition function.  Second, our method allows us to
perform the calculation with a deterministic matrix added to the Dirac
matrix.  This is not easily possible in the standard
orthogonal-polynomial method because this approach rests on the
rotation invariance of the matrix ensemble which is fully broken due
to the presence of the deterministic offset.

We point out again that the microscopic correlation functions
\eqref{final} computed in RMT are universal in the sense that they are
expected to agree with the microscopic spectral correlations of the
Dirac operator in full QCD.  We hope to be able to compare the random
matrix results with data from lattice gauge simulations with dynamical
fermions in the near future.

\section*{Acknowledgments}

We thank H.A. Weidenm\"uller for discussions.  This work was supported
in part by DFG project We 655/15-1 (TW) and by the Heisenberg
foundation (TG).

\appendix

\section{Derivation of Eq.~(\protect\ref{eq1.7a})}
\label{app:Zktilde}

To show that Eq.~\eqref{eq1.7a} holds, we write the matrix $W$ in
spherical coordinates, $W=U\Lambda\bar V$ with $U\in\mathrm{U}(N)$,
$\bar V\in\mathrm{U}(N)/\mathrm{U}^N(1)$, and
$\Lambda=\diag(\lambda_1,\dots,\lambda_N)$, where the $\lambda_n$ are
real and nonnegative \cite{Guhr96}.  The integration measure
transforms according to
\begin{equation}
  d[W]=J(\Lambda)d[\Lambda]d\mu(U)d\mu(\bar V)
\end{equation}
with
\begin{equation}
  d[\Lambda]=\prod_{n=1}^N d\lambda_n \qquad \text{and} \qquad
  J(\Lambda)=\Delta_N^2(\Lambda^2)\prod_{n=1}^N \lambda_n\:.
\end{equation}
The integrations over the $\lambda_n$ extend from 0 to $\infty$,
$d\mu(U)$ and $d\mu(\bar V)$ are the invariant Haar measures, and the
Vandermonde determinant is defined in Eq.~\eqref{berezinian}.  In the
integration over $d[W]$ in Eq.~\eqref{eq1.5}, we shift $W$ and
$W^\dagger$ by $-Y$ and obtain 
\begin{align}
  \label{Zk}
  Z_k(J)&=\frac{1}{\mathcal{N}}\int d[\Lambda]d\mu(U)d\mu(\bar V)
  J(\Lambda)e^{-N\Sigma^2\tr(W-Y)(W^\dagger-Y)} \prod_{f=1}^{N_f}
  \det\begin{bmatrix}im_f&-W\\-W^\dagger&im_f\end{bmatrix}\nonumber\\
  &\qquad\qquad\times
  \prod_{p=1}^k\det\begin{bmatrix}x_p&-W\\-W^\dagger&x_p\end{bmatrix}
  \im\frac{1}{\det\begin{bmatrix}x_p^+-J_p&-W\\
      -W^\dagger&x_p^+-J_p\end{bmatrix}}\nonumber\\
  &=\frac{e^{-N\Sigma^2\tr Y^2}}{\mathcal{N}}\int d[\Lambda] 
  F_N(\Lambda)G_N(\Lambda)H_N(\Lambda)
\end{align}
with
\begin{align}
  F_N(\Lambda)&=\Delta_N^2(\Lambda^2)\prod_{n=1}^N
  \lambda_n e^{-N\Sigma^2\lambda_n^2}
  \prod_{f=1}^{N_f}(-m_f^2-\lambda_n^2)\:,\\
  G_N(\Lambda)&=\prod_{p=1}^k \im\prod_{n=1}^N
  \frac{x_p^2-\lambda_n^2}{(x_p^+-J_p)^2-\lambda_n^2}\:,\\
  \label{HNdef}
  H_N(\Lambda)&=\int d\mu(U)d\mu(\bar V)
  e^{N\Sigma^2\tr(W+W^\dagger)Y}\:.
\end{align}
Analogously, we obtain for Eq.~\eqref{eq1.7}
\begin{equation}
  \label{Zktilde}
  \tilde Z_k(J)=\frac{e^{-N\Sigma^2\tr Y^2}}{\tilde\mathcal{N}}
  \int d[\Lambda]F_N(\Lambda)G_N(\Lambda)H_N(\Lambda) 
  \prod_{f=1}^{N_f}\im\prod_{n=1}^N\frac{1}{(a_f^+)^2-\lambda_n^2}\:.
\end{equation}
The normalization factors $\mathcal{N}$ and $\tilde\mathcal{N}$ follow
by setting $G_N(\Lambda)$ to unity in Eqs.~\eqref{Zk} and
\eqref{Zktilde}, respectively.  Throughout the remainder of this
section, we assume that the dummy variables $a_f$ are pairwise
different.

We proceed by converting the product over $n$ in \eqref{Zktilde} to a
sum,
\begin{equation}
  \label{sum}
  \prod_{n=1}^N\frac{1}{(a_f^+)^2-\lambda_n^2}=
  \frac{1}{\Delta_N(\Lambda^2)}\sum_{n=1}^N(-1)^{N-n}
  \frac{\Delta_{N-1}(\Lambda_{(n)}^2)}{(a_f^+)^2-\lambda_n^2}\:,
\end{equation}
where the subscript $(n)$ means that $\lambda_n$ is omitted in
$\Lambda$.  We now have
\begin{align}
  \label{delta}
  \im\frac{1}{(a_f^+)^2-\lambda_n^2}&=-\frac{\pi}{2a_f}\left[
    \delta(\lambda_n-a_f)+\delta(\lambda_n+a_f)\right]\nonumber\\
  &\longrightarrow -\frac{\pi}{2a_f}\,\delta(\lambda_n-a_f)\:,
\end{align}
where in the last step we have used the fact that the integrations
over the $\lambda_n$ extend from 0 to $\infty$ and that $a_f>0$ (since
we consider the limit $a_f\to\infty$).  Thus, out of the $N$
integrations over the $\lambda_n$ in Eq.~\eqref{Zktilde}, $N_f$ can be
done using the $\delta$-functions.  Using the symmetry of the
integrand with respect to the labeling of the $\lambda_n$, we can
choose to integrate over the last $N_f$ variables,
$\lambda_{N-N_f+1},\dots,\lambda_N$, by relabeling the $\lambda_n$
appropriately and multiplying by a combinatorial factor.  (The
functions $F_N(\Lambda)$ and $G_N(\Lambda)$ are obviously symmetric
under interchanges $\lambda_n \leftrightarrow \lambda_m$, for
$H_N(\Lambda)$ this follows from Eq.~\eqref{HN} below.)

We briefly pause to explain the essence of the proof of
Eq.~\eqref{eq1.7a}.  After performing the $N_f$ integrations in
Eq.~\eqref{Zktilde} and taking the limit $\{a_f\}\to\infty$, one
obtains a result which is essentially equal to the expression for
$Z_k(J)$ in Eq.~\eqref{Zk}, the only difference being that the
integration is over $N-N_f$ variables $\lambda_n$ instead of over $N$
variables.  In the limit $N\to\infty$, this difference can be
neglected.  (There are some additional prefactors which will be
canceled by identical factors in the normalization
$\tilde\mathcal{N}$.)

By performing the $N_f$ integrations in Eq.~\eqref{Zktilde} using the
$\delta$-functions of Eq.\ \eqref{delta} the variables
$\lambda_{N-N_f+1},\dots,\lambda_N$ are replaced by
$a_1,\dots,a_{N_f}$, respectively.  We define
$\Lambda'=\diag(\lambda_1,\dots,\lambda_{N-N_f})$ and
$\bar\Lambda=\diag(\lambda_1,\dots,\lambda_{N-N_f},a_1,\dots,a_{N_f})$.
The various contributions in the integrand of \eqref{Zktilde} become
\begin{align}
  d[\Lambda]&\longrightarrow d[\Lambda']\:,\\
  F_N(\Lambda)&\longrightarrow F_{N-N_f}(\Lambda')\Delta_{N_f}^2(a^2)
  \prod_{f=1}^{N_f}a_fe^{-N\Sigma^2a_f^2}\prod_{n=1}^{N-N_f}
  (a_f^2-\lambda_n^2)^2\prod_{f'=1}^{N_f}(-m_{f'}^2-a_f^2)\nonumber\\
  &\longrightarrow \mathcal{C}_1(a,m)F_{N-N_f}(\Lambda') 
  \quad\text{as $\{a_f\}\to\infty$}\:,\\
  G_N(\Lambda)&\longrightarrow \prod_{p=1}^k \im\prod_{n=1}^{N-N_f}
  \frac{x_p^2-\lambda_n^2}{(x_p^+-J_p)^2-\lambda_n^2}
  \prod_{f=1}^{N_f}\frac{x_p^2-a_f^2}{(x_p^+-J_p)^2-a_f^2}\nonumber\\
  &\longrightarrow G_{N-N_f}(\Lambda') \quad\text{as
  $\{a_f\}\to\infty$}\:. 
\end{align}
Here, $\mathcal{C}_1(a,m)$ is a function which no longer depends on the
$\lambda_n$.  {}From the product over the imaginary parts in
Eq.~\eqref{Zktilde} we obtain with Eqs.~\eqref{sum} and \eqref{delta}
and after appropriate relabeling of the $\lambda_n$
\begin{align}
  \int_0^\infty & d\lambda_{N-N_f+1}\cdots d\lambda_{N}
  \prod_{f=1}^{N_f}\im\prod_{n=1}^N\frac{1}{(a_f^+)^2-\lambda_n^2}
  \nonumber\\
  &\longrightarrow \frac{1}{\Delta_N^{N_f}(\bar\Lambda^2)}
  \frac{(-\pi/2)^{N_f}}{a_1\cdots a_{N_f}}\frac{N!}{(N-N_f)!}
  \prod_{f=1}^{N_f}(-1)^{N_f-f}
  \Delta_{N-1}(\bar\Lambda_{(N-N_f+f)}^2)\nonumber\\
  &\longrightarrow \mathcal{C}_2(a) \quad\text{as $\{a_f\}\to\infty$}\:,
\end{align}
where $\mathcal{C}_2(a)$ is a function which depends only on the
$a_f$.   

Consider now the angular integrals in Eq.~\eqref{HNdef}.  Using
Eq.~(2.3) of Ref.~\cite{Guhr96}, we have
\begin{equation}
  \label{HN}
  H_N(\Lambda)=
  c\:\frac{\det[I_0(\lambda_n \tilde y_m)]_{n,m=1,\dots,N}}
  {\Delta_N(\Lambda^2)\Delta_N(Y^2)}\:,
\end{equation}
where we have defined $\tilde y_m=2N\Sigma^2y_m$.  The constant $c$
depends neither on $\Lambda$ nor on $Y$.  After the integration over
$\lambda_{N-N_f+1},\dots,\lambda_N$, $\Lambda$ is replaced by
$\bar\Lambda$, i.e., the last $N_f$ entries are $a_1,\dots,a_{N_f}$.
We now expand the determinant in the numerator of Eq.~\eqref{HN} with
respect to the last row.  Without loss of generality we can assume
that $y_1<y_2<\ldots<y_N$.  Using the asymptotic behavior of $I_0$,
only the term proportional to $I_0(a_{N_f}\tilde y_N)$ remains in the
limit $a_{N_f}\to\infty$.  The other terms are suppressed by factors
of $\exp(-a_{N_f}(\tilde y_N-\tilde y_n))$, where $n=1,\dots,N-1$.
Using the same argument for the remaining $N_f-1$ bottom rows, we
obtain for $\{a_f\}\to\infty$
\begin{align}
  \det[I_0(\bar\lambda_n \tilde y_m)]_{n,m=1,\dots,N}\longrightarrow &
  \det[I_0(\lambda_n \tilde y_m)]_{n,m=1,\dots,N-N_f}\nonumber\\
  &\times\det[I_0(a_f\tilde y_{N-N_f+g})]_{f,g=1,\dots,N_f}\:.
\end{align}
Denoting $Y'=\diag(y_1,\dots,y_{N-N_f})$ and
$Y''=\diag(y_{N-N_f+1},\dots,y_N)$, we have
\begin{align}
  \Delta_N(\bar\Lambda^2)&=\Delta_{N-N_f}({\Lambda'}^2)\Delta_{N_f}(a^2)
  \prod_{f=1}^{N_f}\prod_{n=1}^{N-N_f}(a_f^2-\lambda_n^2)\:,\\
  \Delta_N(Y^2)&=\Delta_{N-N_f}({Y'}^2)\Delta_{N_f}({Y''}^2)
  \prod_{f=1}^{N_f}\prod_{n=1}^{N-N_f}(y_{N-N_f+f}^2-y_n^2)
\end{align}
and thus obtain
\begin{equation}
  H_N(\Lambda)\longrightarrow \mathcal{C}_3(a,Y) H_{N-N_f}(\Lambda')
  \quad\text{as $\{a_f\}\to\infty$}\:. 
\end{equation}
The function $\mathcal{C}_3(a,Y)$ no longer depends on the
$\lambda_n$.  Note that the last $N_f$ entries of the diagonal matrix
$Y$ have effectively disappeared from the problem.  However, this
effect is negligible in the limit $N\to\infty$.

Collecting the various terms, we finally obtain for $\{a_f\}\to\infty$
\begin{align}
  \tilde Z_k(J)=&\frac{1}{\tilde\mathcal{N}}\,e^{-N\Sigma^2\tr Y^2}\,
  \mathcal{C}_1(a,m)\mathcal{C}_2(a)\mathcal{C}_3(a,Y)\nonumber\\
  &\times\int d[\Lambda']F_{N-N_f}(\Lambda')G_{N-N_f}(\Lambda')
  H_{N-N_f}(\Lambda')\:.
\end{align}
The normalization factor $\tilde\mathcal{N}$ is obtained by setting
$G_{N-N_f}(\Lambda')$ to unity so that
\begin{equation}
  \lim_{\{a_f\}\to\infty}\tilde Z_k(J)=
  \frac{\displaystyle{\int d[\Lambda']F_{N-N_f}(\Lambda')
    G_{N-N_f}(\Lambda')H_{N-N_f}(\Lambda')}}
  {\displaystyle{\int d[\Lambda']F_{N-N_f}(\Lambda')H_{N-N_f}(\Lambda')}}
\end{equation}
which is equal to $Z_k(J)$ in Eq.~\eqref{Zk} with $N$ replaced by
$N-N_f$.  In the limit $N\to\infty$, this difference is negligible.
This completes the proof.

\section{Derivation of Eq.~(\protect\ref{eq3.5})}
\label{app:limit} 

We wish to compute the $\{\alpha_f\}\to\infty$ limit in
Eq.~\eqref{eq3.3}.  It is convenient to divide both numerator and
denominator by the Vandermonde determinant $\Delta_{N_f}(\alpha)$, see
Eq.~\eqref{berezinian}, and to compute the $\{\alpha_f\}\to\infty$
limit of the quantity
\begin{equation}
  \mathcal{R}=
  \frac{\det[C(\tilde z_p,\tilde \zeta_q)]_{p,q=1,\dots,k+N_f}}
  {\Delta_{N_f}(\alpha)}
\end{equation}
with a kernel $C$ given in Eq.~\eqref{kernel}, $\tilde z_p$ and
$\tilde\zeta_q$ defined after Eq.~\eqref{kernel}, and $z_p$ and $\zeta_q$
given in Eqs.~\eqref{z_p} and \eqref{zeta_q}, respectively.  The
denominator in Eq.~\eqref{eq3.3} then follows immediately by setting
$k=0$.  The $\alpha$-dependence in the numerator determinant of
$\mathcal{R}$ is found in the last $N_f$ rows,
\begin{equation}
  \mathcal{R}=\frac{1}{\Delta_{N_f}(\alpha)}
  \begin{vmatrix}
    \hdotsfor[1.5]{6}\\
    C(\alpha_1,u_1)&\cdots&C(\alpha_1,u_k)&
    C(\alpha_1,i\mu_1)&\cdots&C(\alpha_1,i\mu_{N_f})\\
    \vdots&&\vdots&\vdots&&\vdots\\
    C(\alpha_{N_f},u_1)&\cdots&C(\alpha_{N_f},u_k)&
    C(\alpha_{N_f},i\mu_1)&\cdots&C(\alpha_{N_f},i\mu_{N_f})
    \nonumber
  \end{vmatrix}\:,
\end{equation}
or, indicating rows and their $\alpha_i$-dependence schematically
by $r(\alpha_i)$,
\begin{equation}
  \mathcal{R}=\frac{1}{\Delta_{N_f}(\alpha)}
  \begin{vmatrix}
    \hdotsfor[1.5]{3}\\
    \cdots&r(\alpha_1)&\cdots\\
    &\vdots& \\
    \cdots&r(\alpha_{N_f})&\cdots
  \end{vmatrix}\:.
\end{equation}
Before taking the $\{\alpha_f\}\to\infty$ limit, we need to take the
limits $\alpha_f\to\alpha_g$ for all $f<g$.  Let us start with
$\alpha_1$.  We subtract the row $r(\alpha_1)$ from all following
rows, $r(\alpha_2)$ to $r(\alpha_{N_f})$, without changing the value
of the determinant. The Vandermonde determinant supplies factors of
the kind $1/(\alpha_i-\alpha_1)$ for $2\le i\le N_f$.  Thus, for the
$i$-th row $r(\alpha_i)$, we arrive at
\begin{equation}
  r(\alpha_i)\rightarrow r(\alpha_i)-r(\alpha_1)
  \rightarrow \frac{r(\alpha_i)-r(\alpha_1)}{\alpha_i-\alpha_1}
  \xrightarrow{\alpha_1 \rightarrow \alpha_i}
  \partial_{\alpha_i} r(\alpha_i)\:.
\end{equation}
Next, the resulting second row $\partial_{\alpha_2} r(\alpha_2)$ is
subtracted from all following rows, $\partial_{\alpha_3}r(\alpha_3)$
to $\partial_{\alpha_{N_f}}r(\alpha_{N_f})$.  Including factors
supplied by the Vandermonde determinant, the $j$-th row $(3 \leq j
\leq N_f)$ thus becomes
\begin{equation}
  \partial_{\alpha_j}r(\alpha_j)\rightarrow
  \partial_{\alpha_j}r(\alpha_j)-\partial_{\alpha_2}r(\alpha_2)
  \rightarrow
  \frac{\partial_{\alpha_j}r(\alpha_j)-\partial_{\alpha_2}r(\alpha_2)}
  {\alpha_j-\alpha_2}
  \xrightarrow{\alpha_2 \rightarrow \alpha_j}
  \partial_{\alpha_j}^2 r(\alpha_j)
\end{equation}
Analogous steps for the subsequent rows lead to higher derivatives, at
the same time eating up all factors contained in the Vandermonde
determinant.  Finally, we obtain
\begin{align}
  \label{limR}
  \lim_{\{\alpha_f\}\to\infty}\mathcal{R}=\lim_{\alpha\to\infty}
  \begin{vmatrix}
    \hdotsfor[2]{3}\\
    \dots&r(\alpha)&\dots\\
    \dots& \partial_{\alpha}r(\alpha)&\dots\\
    &\vdots&\\
    \dots& \partial_{\alpha}^{N_f-1}r(\alpha)&\dots
  \end{vmatrix}\:,
\end{align}
where $\alpha$ is now a simple number.

We now consider a generic derivative appearing in \eqref{limR}.
Recall that $r(\alpha)$ stands for $C(\alpha,x)$, where $x$ can be
one of the $u_p$ or one of the $i\mu_f$.  The following manipulations
will become more transparent by considering the first few special
cases.  We have
\begin{equation}
  \label{d0}
  \partial^0_{\alpha} C(\alpha,x)=C(\alpha,x)
  \xrightarrow{\alpha\to\infty}\frac1\alpha J_1(\alpha)J_0(x)\:.
\end{equation}
The point is that this result is proportional to $J_0(x)$ in the
large-$\alpha$ limit.  The first derivative becomes
\begin{align}
  \label{d1}
  \partial^1_{\alpha} C(\alpha,x)\xrightarrow{\alpha\to\infty}&
  \left[\frac1\alpha J_0(\alpha)-\frac{2}{\alpha^2}J_1(\alpha)\right]
  J_0(x)+\frac{1}{\alpha^2}J_0(\alpha)xJ_1(x)\nonumber\\
  \xrightarrow{\phantom{\alpha\to\infty}}&
  \:\frac{1}{\alpha^2}J_0(\alpha)xJ_1(x)\:,
\end{align}
where the term proportional to $J_0(x)$ has been eliminated by
subtracting an appropriate multiple of the row \eqref{d0}.  Thus, the
row containing the first derivative is proportional to $xJ_1(x)$ in
the large-$\alpha$ limit.  For the second derivative, we proceed
analogously and obtain, in the large-$\alpha$ limit, terms
proportional to $J_0(x)$, $xJ_1(x)$, and $x^2J_0(x)$.  The first two
of these terms can be eliminated by subtracting appropriate multiples
of the rows \eqref{d0} and \eqref{d1}, respectively.  Thus, the row
containing the second derivative is proportional to $x^2J_0(x)$ in the
large-$\alpha$ limit.  We now proceed to a general $m-$fold derivative
for which it is useful to expand the denominator of $C$ in a geometric
series, leading to
\begin{align}
  \label{app1.2}
  \partial^m_{\alpha} C(\alpha,x)=&
  \sum_{l=0}^m \binom{m}{l}
  \partial_{\alpha}^l\left(\frac{1}{\alpha^2-x^2} \right)
  \partial_{\alpha}^{m-l}\left[
    \alpha J_1(\alpha)J_0(x)-xJ_0(\alpha)J_1(x)\right]\nonumber\\
  =&\sum_{l=0}^m \binom{m}{l}\partial_\alpha^l\left(
  \sum_{n=1}^{\infty}\frac{x^{2n-2}}{\alpha^{2n}}\right)
  \partial_{\alpha}^{m-l}\left[
    \alpha J_1(\alpha)J_0(x)-xJ_0(\alpha)J_1(x)\right]\nonumber\\
  =&\sum_{l=0}^m \binom{m}{l}\sum_{n=1}^{\infty}(-1)^l
  \frac{(2n+l-1)!}{(2n-1)!}\frac{x^{2n-2}}{\alpha^{2n+l}}
  \nonumber\\
  &\!\!\times\left[\left(\alpha J_1^{(m-l)}(\alpha)
      +(m-l)J_1^{(m-l-1)}(\alpha) \right)J_0(x) 
    - xJ_0^{(m-l)}(\alpha)J_1(x)\right] .
\end{align}
It will not be necessary to perform the derivatives of $J_0(\alpha)$
and $J_1(\alpha)$.  The expression \eqref{app1.2} contains terms
proportional to $J_0(x)$, $xJ_1(x)$, $\dots$ up to $x^mJ_0(x)$ (if $m$
is even) or $x^mJ_1(x)$ (if $m$ is odd).  All higher order terms in
$x$ are suppressed by powers of $\alpha$ in the large-$\alpha$ limit,
see the sum over $n$.  Subtracting appropriate multiples of previous
rows, only the term proportional to $x^mJ_{m\,\text{mod}\,2}(x)$
remains.  By using Bessel function recursion relations and adding
appropriate multiples of previous rows, $J_{m\,\text{mod}\,2}(x)$ can
be replaced by $J_m(x)$.  We thus obtain
\begin{equation}
  \partial^m_{\alpha} C(\alpha,x) \xrightarrow{\alpha\to\infty}
  x^mJ_m(x)  \times f_m(\alpha)
\end{equation}
with unspecified functions $f_m(\alpha)$.  Defining
\begin{equation}
  \label{F}
  \mathcal{F}=\lim_{\alpha\to\infty}\prod_{m=0}^{N_f-1}f_m(\alpha)
\end{equation}
and noting that $(i\mu)^mJ_m(i\mu)=(-\mu)^mI_m(\mu)$, we arrive at
\begin{equation}
  \label{numlimit}
  \lim_{\{\alpha_f\}\to\infty}
  \frac{\det[C(\tilde z_p,\tilde \zeta_q)]_{p,q=1,\dots,k+N_f}}
  {\Delta_{N_f}(\alpha)} = \mathcal{F}\cdot\det\mathcal{A}
\end{equation}
with $\mathcal{A}$ given in Eq.~\eqref{Amatrix}.  Setting $k=0$, we
obtain
\begin{equation}
  \label{denlimit}
  \lim_{\{\alpha_f\}\to\infty}
  \frac{\det[C(\alpha_f,i\mu_g)]_{f,g=1,\dots,N_f}}
  {\Delta_{N_f}(\alpha)} = \mathcal{F}\cdot\det\mathcal{B}
\end{equation}
with $\mathcal{B}$ given in Eq.~\eqref{Bmatrix}.  Taking the ratio of
\eqref{numlimit} and \eqref{denlimit}, we finally arrive at
\eqref{eq3.5}.  Note that it is not necessary to evaluate \eqref{F}
since $\mathcal{F}$ drops out of the final result.

\end{document}